\documentclass[aps,prl,twocolumn,superscriptaddress]{revtex4}
\usepackage{graphicx}
\usepackage{bm}
\newcommand{\fig}[1]   {Fig.~\ref{#1}} 
\newcommand{\etal}	{\textit{et~al.}}
\begin{document}
\title{A quantitative study of quasiparticle traps using the single-Cooper-pair-transistor}
\author{N. A. Court}
\email{ncourt@phys.unsw.edu.au}
\author{A. J. Ferguson}
\altaffiliation[Present Address: ]{Cavendish Laboratory, University of Cambridge, J.J. Thomson Avenue, CB3 0HE, UK.}
\affiliation{Australian Research Council Centre of Excellence for Quantum Computer Technology, University of New South Wales, Sydney NSW 2052, Australia.}%
\author{Roman Lutchyn}
\affiliation{Joint Quantum Institute, University of Maryland, College Park, MD 20742.}
\author{R. G. Clark}%
\affiliation{Australian Research Council Centre of Excellence for Quantum Computer Technology, University of New South Wales, Sydney NSW 2052, Australia.}%
\date{\today}

\begin{abstract}
We use radio-frequency reflectometry to measure quasiparticle tunneling rates in the single-Cooper-pair-transistor. Devices with and without quasiparticle traps in proximity to the island are studied. A $10^2$ to $10^3$-fold reduction in the quasiparticle tunneling rate onto the island is observed in the case of quasiparticle traps. In the quasiparticle trap samples we also measure a commensurate decrease in quasiparticle tunneling rate off the island.
\end{abstract}

\maketitle
In a superconductor charge may be transported coherently by Cooper-pairs and incoherently by quasiparticles. For nanoscale Coulomb blockade devices such as the single-Cooper-pair-transistor (SCPT) and Cooper-pair-box (CPB), these two modes of transport are mutually exclusive \cite{joyez_prl_94,aumentado_prl_04,ferguson_prl_06,naaman_prb_06}. Coherent charge transport is temporarily halted while a quasiparticle is present on the device island. This effect, caused by the electrostatic energy cost of the quasiparticle, is known as quasiparticle poisoning and leads to Cooper-pair tunneling being stochastically switched on and off.  Quasiparticle poisoning is usually unwanted, especially in charge-based superconducting qubits (e.g. the CPB) where it changes the qubit bias point in a random telegraphic way and places limits on quantum coherence \cite{lang_ieee_03,duty_prb_04,lutchyn_prb_05}.

In order to suppress quasiparticle poisoning the density of quasiparticles in the device leads should be reduced. In principle this can be simply achieved by lowering the temperature, however this is not always sufficient and an additional approach is to use so-called ``quasiparticle traps''. These are normal metal regions contacting the superconducting leads in proximity to the device island \cite{joyez_prl_94}. In this paper we measure quasiparticle tunneling rates for SCPTs with and without quasiparticle traps. We report quantitatively on the effect of the traps, and discuss the implications of our results for designing superconducting Coulomb blockade devices with reduced quasiparticle poisoning.

Quasiparticle traps provide a sink to reduce quasiparticle density. When a superconductor is brought in contact with a normal metal, an induced superconducting gap develops in the normal metal, this is known as the proximity effect. Quasiparticles in the superconductor diffuse into the normal metal side where they relax towards the gap edge on a timescale which depends on the electron-phonon coupling \cite{goldie_prl_90,dekorte_procspie_92}. The trapped quasiparticles eventually recombine, emitting a phonon with insufficient energy to break a Cooper pair in the superconductor. Quasiparticle traps have been studied in the context of x-ray photon detection \cite{goldie_prl_90,nahum_apl_95} and cooling through superconductor-normal junctions \cite{ullom_prb_2000,pekola_apl_2000}. Joyez \textit{et~al.} first applied quasiparticle traps to reduce quasiparticle poisoning in SCPTs \cite{joyez_prl_94}, copper leads were contacted at a distance of $<$1 $\mu$m from the junctions and a 2$e$-periodic supercurrent was observed at low temperatures. Quasiparticle traps have often been used where quantum coherence is of interest \cite{bouchiat_pscr_98,lang_ieee_03,chiorescu_nature_04,ithier_prb_05,bladh_njp_05}. A variety of results have been achieved with immunity from quasiparticle poisoning not always evident \cite{schniderman_condmat_07, bladh_njp_05}: therefore providing a motivation for this work.

\begin{figure}[b]
\begin{center}
\includegraphics[width=6.5cm]{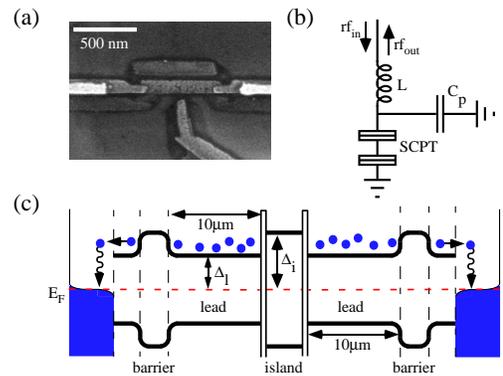}
\end{center}
\caption{(a) Micrograph of a QT device with an island volume $vol_i$=750 nm$\times$125 nm$\times$7 nm. Bright features are the normal metal traps 200-300 nm from SCPT junctions. (b) The rf-circuit has a resonance frequency of $\sim$ 320 MHz and includes a chip inductor ($L =$ 470 nH), a parasitic capacitance ($C_p =$ 0.53 pF) and the SCPT. (c) Energy gap profile showing isolation of the leads from the normal metal bond-pads.}\label{fig1trap}
\end{figure}

We make use of the enhanced superconducting gap of aluminum with decreasing film thickness \cite{court_condmat_07}. The SCPT island gap is made greater than the lead gap ($\Delta_i>\Delta_l$) to reduce the depth of the quasiparticle potential well on the island \cite{aumentado_prl_04}. Also, the leads of the SCPTs are isolated from the normal metal contacts by a region of higher superconducting gap, forming a quasiparticle barrier. This was designed to reduce the effect of quasiparticles in the leads being trapped by the normal metal bond-pads, allowing the full effect of the intentional traps to be observed [\fig{fig1trap}(c)]. The island and quasiparticle barrier layers are 7 nm thick ($\Delta_i =$  298 $\pm$ 9.4 $\mu$eV) while the leads are 30 nm thick ($\Delta_l=$ 209  $\pm$ 11 $\mu$eV) \cite{court_condmat_07}. Devices were fabricated with and without direct contact to quasiparticle traps. We refer to these device types as QT (quasiparticle traps) and NT (no traps) respectively [\fig{fig1trap}(a)]. AuPd alloy was chosen as the trap metal as we find this reliably gives a low contact resistance to aluminum. Our quasiparticle traps were typically 200-300 nm from the SCPT tunnel junctions and have approximately the same volume as the leads ($vol =$ 2$\times$10 $\mu$m$\times$100 nm$\times$30 nm).

\begin{figure}[t]
\begin{center}
\includegraphics[width=6.5cm]{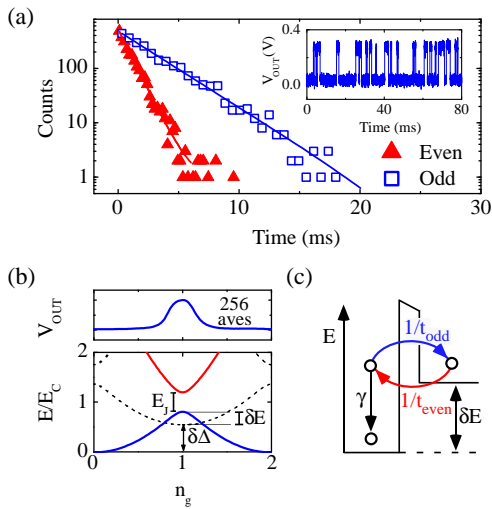}
\end{center}
\caption{(a) Histogram of times spent in the even and odd states for a QT device, solid lines are fitted exponentials. Here $t_{odd}=$ 3.1 ms and $t_{even}=$ 964 $\mu$s. Inset: switching between even and odd states at $n_g =$ 1. (b) Upper: Averaged supercurrent oscillations and Lower: energy band diagram of the SCPT at zero current bias. The quasiparticle bands (dashed line) are lifted by an energy $\delta\Delta$ creating a potential well of energy $\delta{E}$ relative to the lower Cooper pair band (solid line). (c) Allowed transitions of a quasiparticle including the relaxation rate to the bottom of the potential well on the island.}\label{fig2trap}
\end{figure}

The SCPT bandstructure determines the energetics of quasiparticle tunneling [\fig{fig2trap}(b)]. The difference in superconducting gaps between the leads and the island lifts the energy minima of the quasiparticle state by an energy $\delta\Delta = \Delta_i - \Delta_l$ \cite{aumentado_prl_04}. In our devices, the quasiparticle state remains lower in energy than the ground band of the SCPT at $n_g =$ 1. Therefore a quasiparticle potential well is formed on the SCPT island with a depth given by $\delta{E} = E_c - \frac{E_{J}}{2} - \delta\Delta$ \cite{aumentado_prl_04}. The presence of a well means that a quasiparticle in the leads can tunnel into the island, relax, and after some time can be thermally activated out. We refer to the time constants for the poisoning process as $t_{even}$ (the even state lifetime) and the unpoisoning process as $t_{odd}$ (the odd state lifetime). From normal state Coulomb diamonds ($B =$ 3T) we find the charging energy of the principal QT (NT) device as, $E_c = e^2/2C_{\Sigma} =$ 170 $\mu$eV (185 $\mu$eV); the total device resistance at 4.2 K is  42 k$\Omega$ (54 k$\Omega$); and the Josephson energy per junction from the Ambegoakar-Baratoff relation is $E_J =$ 37 $\mu$eV (28 $\mu$eV). The $\delta\Delta$ we expect from SIS junction measurements is 89 $\pm$ 12 $\mu$eV. Hence, the expected depths of the island potential well are $\delta{E} =$ 62.5 $\mu$eV  (82 $\mu$eV), in both cases $\delta{E}\gg{kT}$. For our set of devices $E_c$ and $E_J$ are consistent to within 25$\%$.

Radio-frequency reflectometry is employed to perform high-bandwidth measurements of quasiparticle tunneling \cite{schoelkopf_sci_1998}. This consists of embedding the SCPT in a resonant circuit and detecting the amplitude (and phase) of a small rf signal reflected from the circuit at resonance [\fig{fig1trap}(b)]. The reflected carrier signal passes through a low noise cryogenic amplifier before being further amplified and demodulated. Briefly, the presence of a quasiparticle on the device island shifts the supercurrent oscillation by a gate charge of $e$. This changes the impedance of the SCPT and changes the amplitude and phase of the reflection coefficient \cite{ferguson_prl_06, naaman_prb_06}.

We monitor the demodulated signal rf signal ($V_{OUT}$) with the device biased on a supercurrent peak (upper graph \fig{fig2trap}(b)) and observe two-level switching due to quasiparticle tunneling [\fig{fig2trap}(a) inset]. Traces consist of $10^6$ data points and $\sim1\times10^4$ switching events, and are analyzed by comparing $V_{OUT}$ to a median value. We plot a histogram of times spent in each state [\fig{fig2trap}(a)], an exponential is fitted to the histogram and the decay constant defines the state lifetimes. A good fit indicates Poissonian tunneling processes. Finite receiver bandwidth causes a systematic overestimate of time constants, and we follow \cite{naaman_prl_06} to correct for this.  Low pass filters of between 1.0 and 10.7 MHz are used and the receiver bandwidth is estimated for each filter. We operate at an rf-bias where the time constants are not significantly perturbed, corresponding to a carrier power of -100.5 dBm (-104 dBm) for the QT (NT) devices \cite{ferguson_prl_06}.

\begin{figure}[t]
\begin{center}
\includegraphics[width=6.8cm]{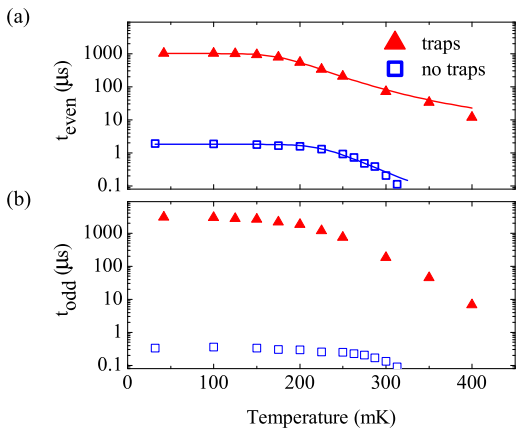}
\end{center}
\caption{ Temperature dependence of (a) $t_{even}$  and (b) $t_{odd}$ for a QT device (triangles) and a NT device (squares). Solid lines are fits to the quasiparticle density in the leads (see text).} \label{fig3trap}
\end{figure}

For the QT device a dramatic increase in $t_{even}$ (reduction in poisoning rate), compared to the NT devices, is observed across the whole temperature range [\fig{fig3trap}(a)]. The saturated low temperature values are $t_{even}=$ 1 ms for the QT case and $t_{even}=$ 1.9 $\mu$s for the NT case. The equilibrium expression for quasiparticle density is $n_{qp}=D(\epsilon_{F})\sqrt{2\pi\Delta_l kT}$ exp$(-\Delta_l/kT)$, where $D(\epsilon_{F}) =$ 1.45$\times$10$^{47}$ $m^{-3}J^{-1}$ is the normal-state density of states (including spin) of aluminum at the Fermi energy. Since $t_{even}^{-1}$ is proportional to the quasiparticle density in the leads, we fit to the following expression $t_{even}^{-1}= B\left(n_{qp}(T_{qp},\Delta_l) + n_{qp}(T,\Delta_l)\right)$, where $B$ is a constant, $T_{qp}$ is the saturation temperature and $n_{qp}$ refers to the quasiparticle density in the leads \cite{ferguson_prl_06}. For the NT and QT devices, we find the minimum quasiparticle temperatures are $T_{qp} =$ 248 $\pm$ 3 mK and $T_{qp} =$ 204 $\pm$ 1 mK; the superconducting gaps $\Delta_l =$ 210 $\pm$ 20 $\mu$eV and $\Delta_l =$ 120 $\pm$ 3 $\mu$eV; and the constant $B =$ 2.2 $\pm$ 1.9$\times10^{-15}$ $m^3s^{-1}$ and $B =$ 4.0 $\pm$ 0.6$\times10^{-19}$ $m^3s^{-1}$.

A reduction in quasiparticle temperature is seen for the QT device, however this only partially explains the change in $t_{even}$. If the effect of the quasiparticle traps was purely to thermalize the leads we would expect the temperature dependencies to overlie one another at high temperature with the QT case saturating at a lower temperature. To illustrate this further we take the expression for $n_{qp}(T)$ and estimate the difference in $t_{even}$ for the observed change in minimum quasiparticle temperature, $n_{qp}($248 mK$)/n_{qp}($204 mK$)=t_{even}($204 mK$)/t_{even}($248 mK$)\sim$ 9: a relatively small fraction of the total change.

The NT gap parameter extracted for the leads ($\Delta_l =$ 210 $\pm$ 20 $\mu$eV) agrees with our expected value, but a significant suppression is seen for the QT device ($\Delta_l =$ 120 $\pm$ 3 $\mu$eV). We check if this corresponds to a real reduction in $\Delta_l$ by performing a voltage-biased measurement of quasiparticle tunneling threshold. This threshold occurs at $V_{ds}=$2$e\Delta_l +$2$e\Delta_i=$ 998 $\mu$eV for the QT device so, assuming an unchanged $e\Delta_i=$298 $\mu$eV then $e\Delta_l=$ 201 $\mu$eV which is close to the NT case. We conclude that that $\Delta_l$ is not significantly changed by the presence of the normal metal. However, it is not the only quasiparticle activation energy in a superconductor-normal bilayer.

\begin{figure}[t]
\begin{center}
\includegraphics[width=7.4cm]{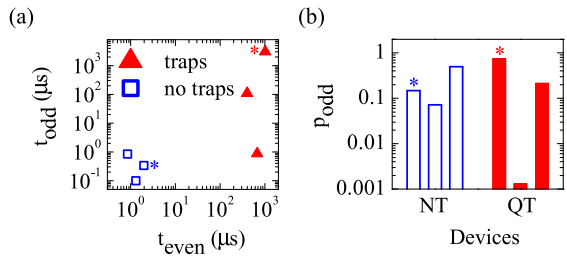}
\end{center}
\caption{ (a) Values of $t_{odd}$ and $t_{even}$ plotted for QT (triangles) and NT devices (squares) at fridge base temperature. (b) The probability of having a quasiparticle on the island ($p_{odd}$), determined from the time constants at this temperature. QT(NT) devices are indicated as filled(unfilled). The principal QT(NT) device is noted with \textbf{*}.} \label{fig4trap}
\end{figure}

Furthermore we can look at the values of the proportionality constant, $B$, for the tunneling rate. Theoretical analysis of the tunneling rates show that $t_{even}^{-1}=\frac{G n^l_{qp}}{2e^2{D(\epsilon_F)}} \nu_i(\delta{E}+\Delta_i) \frac{\delta{E}}{\delta{E}+\Delta_i}$ \cite{lutchyn_prb_07,gapengcase}, where $\nu_{i,l}(E_k)=\frac{E_k}{\sqrt{E_k^2-{\Delta_{i,l}^2}}}$ is the quasiparticle density of states and $G=G_1+G_2$, in S, is the total tunnel barrier conductance.  Thus our proportionality constant is predicted to be $B =$ 1.6$\times$10$^{-15}$ $m^3s^{-1}$ showing close agreement with the measured value in the NT case but not in the QT case. From these fitting parameters it appears that the density of quasiparticle excitations, and their rate of tunneling onto the SCPT island, in the NT case is well-described. However, from the QT device, we observe that the quasiparticle density in the superconducting side of a superconductor-normal bilayer is significantly reduced and is not given by the equilibrium expression for quasiparticle density $n_{qp}(T_{eff})$ at an effective temperature $T_{eff}$. We note that the increased $t_{even}$ for the QT devices is reproduced amongst our sample set which consists of 3 of each device type [\fig{fig4trap}(a)].

We now discuss qualitatively how the quasiparticle density is affected by the presence of a quasiparticle trap. In a superconductor the number of quasiparticles ($N_{qp}$) and phonons ($N_{\omega}$) with $E>2\Delta$ are closely coupled and described by the Taylor-Rothwarf equations \cite{rothwarf_prl_67}. In the presence of a quasiparticle trap, it is necessary to add additional terms to describe loss to (-$\Gamma_TN_{qp}$) and gain from ($\Gamma_UN^{T}_{qp}$) the trap \cite{wilson_prb_04}. Quasiparticle population of the trap is given by $N^T_{qp}$. The constants $G$ and $R$ represent the quasiparticle generation and recombination rates in the superconductor. The quasiparticle master equation for the superconducting film follows. Other equations treat the quasiparticle density in the trap ($N^T_{qp}$) and the phonon density in both films.
\begin{equation}
\frac{dN_{qp}}{dt}=2GN_\omega-\frac{RN_{qp}^2}{vol} - \Gamma_{T} N_{qp} +\Gamma_U N^T_{qp}
\end{equation}

To fully explain the $t_{even}$ behavior of the QT samples, the solution of this equation should be fitted to the data in \fig{fig3trap}(a). We do not perform this analysis but note that the significant reduction in quasiparticle density of the superconducting film implies that trapping, rather than recombination becomes the dominant loss term. In this case, where tunneling to the trap is the fastest timescale, the quasiparticle density in the lead is given by detailed balance between the lead and the trap, $\Gamma_T N_{qp}=\Gamma_U N_{qp}^T$. The trapping ($\Gamma_T$) and un-trapping rates ($\Gamma_U$) will depend on the lead-trap interface. Since $N_{qp}$ is smaller for QT devices, it follows that the number of recombination phonons ($N_{\omega}$) is also reduced.

Now we turn to the behavior of $t_{odd}$. Two of three QT samples show a significant increase in $t_{odd}$ over the NT devices \cite{vol}. The increase is comparable to that for $t_{even}$: there is a factor of 9000 for the principal QT and NT devices [\fig{fig3trap}(b)]. We can estimate the reduction in temperature needed from $t^{-1}_{odd}\propto\exp{(-dE/kT)}$. Taking the mean value of quasiparticle potential depth ($dE=72$ $\mu eV$) for the two principal devices, a reduction in phonon temperature from 248 mK to 67 mK can explain the difference observed in $t_{odd}$. Phonons (with $E\geq2\Delta$), resulting from quasiparticle recombination should be considered in addition to the thermal phonons. If the effect of the thermal phonons was to dominate the un-poisoning process, then $t_{odd}$ should saturate at the phonon temperature of the sample. We expect the phonon temperature to be lower than the quasiparticle temperature and close to the base temperature of the dilution refrigerator ($T\sim50$ mK). In these devices $t_{odd}$ saturates at a temperature close to $t_{even}$ implying that recombination phonons may play a significant role in the unpoisoning process. The relative effect of thermal and recombination phonons may depend on sample parameters. In a previous device, with lower charging energy \cite{ferguson_prl_06}, $t_{odd}$ was observed to saturate at a lower temperature that $t_{even}$.

In the QT device with the short $t_{odd} =$ 880 ns [\fig{fig4trap}(a)] we see no temperature dependence of $t_{odd}$, suggesting the quasiparticle exits without thermalization. This behavior may be explained by the existence of two unpoisoning mechanisms \cite{lutchyn_prb_07}. Either the quasiparticle tunnels off the island immediately in an elastic unpoisoning process [\fig{fig2trap}(c)]. Or, it relaxes to the bottom of the potential well with a rate $\gamma$ which strongly depends on the island potential depth ($\gamma\propto dE^{3.5})$. Subsequently thermal activation is required for the unpoisoning event to occur. The escape rate ${t_{odd}}^{-1}=\frac{G\delta_i}{2e^2}\nu(T)\frac{\delta{E}}{\delta{E}+\Delta}$ for the first scenario is governed by the conductance of the junctions, the quasiparticle level spacing on the island, $\delta_i$, and the initial energy of the quasiparticle (of order $T$) \cite{lutchyn_prb_07,gapengcase}. We calculate $t_{odd} =$ 280 ns, in reasonable agreement with the measured value. The reason for the dominance of the elastic process in this sample is not clear, one possibility is variation of the thin-films leading to a reduced potential well depth on the island.

Our measurements are sensitive to unpoisoning and poisoning processes, however this is not always the case. For example, in electrometry of a CPB by a single-electron-transistor, averaged measurements of the box charge are typically performed \cite{duty_prb_04}. This yields the average probability of having a quasiparticle on the island $p_{odd}$ without directly providing details of the individual tunneling rates. Taking our measured times for the QT (NT) device then $p_{odd}=\frac{t_{odd}}{t_{even}+t_{odd}}$ = 0.75 (0.15) at base temperature. The time the SCPT remains poisoned is greater in the QT device, even though the the tunneling rate onto the island is significantly reduced. Thus one might conclude from $p_{odd}$ that the quasiparticle traps actually had a negative effect. Figure 4(b) plots $p_{odd}$ for all the samples measured and shows no obvious trend between the QT and NT devices. This is consistent with the electrometry measurements of a CPB observing no clear benefit of quasiparticle traps \cite{bladh_njp_05}.

In conclusion, quasiparticle traps can definitely help reduce quasiparticle poisoning in superconducting qubits: a reduction of 2-3 orders of magnitude in poisoning rate was observed here. However, quasiparticle traps are only part of the solution to the quasiparticle poisoning problem. They should be used in conjunction with small charging energy devices and graded superconducting gaps to ensure high unpoisoning rates.

The authors would like to thank D. Barber and R. P. Starrett for technical support. This work is supported by the Australian Research Council, the Australian Government, and by the US National Security Agency (NSA) and US Army Research Office (ARO) under Contract No. W911NF-04-1-0290.

\bibliographystyle{apsrev}

\begin{thebibliography}{25}
\bibitem[{gap()}]{gapengcase}
\bibinfo{note}{The expression given in \cite{lutchyn_prb_07} $\Delta_i=\Delta_l$. Here we include a modification
  for different gaps which is valid as long as
  $\Delta_i,\Delta_l>\delta{E}\gg{kT}$.}
\bibitem[{vol()}]{vol} \bibinfo{note}{The island volume for the QT devices was approximately 3 
times that for the NT case. This accounts for a factor of 3 difference in $t_{odd}$ times \cite{lutchyn_prb_07}.}
\end{thebibliography}

\end{document}